\documentclass{elsart}
\usepackage{graphics}
\usepackage{graphicx}
\usepackage{epsfig}
\usepackage{amssymb}
\usepackage[fleqn]{amsmath}
\usepackage{bm}

\begin{document}

\begin{frontmatter}
\title{Chiral dynamics and baryon resonances}

\author[TokyoTech]{Tetsuo Hyodo}

\address[TokyoTech]{Department of Physics, Tokyo Institute of 
Technology, Meguro 152-8551, Japan}

\begin{abstract}
    The structure of baryon resonance in coupled-channel meson-baryon scattering 
    is studied from the viewpoint of chiral dynamics. The meson-baryon scattering
    amplitude can be successfully described together with the properties of the
    resonance in the scattering, by implementing the unitarity condition for the 
    amplitude whose low energy structure is constrained by chiral theorem. 
    Recently, there have been a major progress in the study of the structure of 
    the resonance in chiral dynamics. We review the methods to clarify the 
    structure of the resonance by focusing on the $\Lambda(1405)$ resonance.
\end{abstract}
\end{frontmatter}

\section{INTRODUCTION}

Chiral symmetry is the one of the guiding principles to study the low energy 
hadron physics, since it dictates the dynamics of the Nambu-Goldstone bosons with
hadrons through the chiral low energy theorem. One of the great successes is the 
celebrated Weinberg-Tomozawa theorem~\cite{Weinberg:1966kf,Tomozawa:1966jm}, 
which determines the $\pi N$ and $\pi\pi$ scattering lengths in good agreement 
with experimental data. The current algebra technique was then refined by the 
establishment of power counting~\cite{Weinberg:1979kz}, which enables one to sort
out the effective chiral Lagrangian and amplitude, leading to the systematic 
computation of the higher order corrections in chiral perturbation 
theory~\cite{Gasser:1985gg}. In recent years, the framework of chiral dynamics is
extended to the resonance energy region, with the implementation of the 
coupled-channel unitarity condition of the scattering amplitude. This chiral 
coupled-channel approach has been providing fairly successful description of 
baryon resonances in meson octet-baryon octet 
scattering~\cite{Kaiser:1995eg,Oset:1998it,Oller:2000fj,Lutz:2001yb}. 

Within the phenomenologically successful framework of meson-baryon scattering, it
is interesting to study the structure of the resonances, which reflects the 
nonperturbative dynamics of strong interaction. In general, several constructions
are possible for a baryon resonance, such as three-quark state, meson-baryon 
molecule, and more complicated structures. Physical state may be a superposition 
of all possible components with given quantum numbers, but the clarification of 
the dominant component should help our intuitive understanding of the resonance 
structure. 

Among many resonances, the $\Lambda(1405)$ attracts considerable attention from 
several aspects. There is a long-standing discussion about its internal 
structure; whether it is a conventional three-quark state in quark 
models~\cite{Isgur:1978xj}, or it is composed of meson-baryon 
molecule~\cite{Dalitz:1967fp}. In a recent context, the properties of the
$\Lambda(1405)$ is extensively discussed in relation with the possible 
kaonic nuclei~\cite{Akaishi:2002bg}. Since the $\Lambda(1405)$ lies just below 
the $\bar{K}N$ threshold, it is considered to be a quasi-bound state of 
$\bar{K}N$ 
system, and thus it determines the bare interaction between kaon and nucleon. 
Indeed, it is found that the $\Lambda(1405)$ plays an important role in the 
extrapolation of the $\bar{K}N$ interaction down to far subthreshold energy 
region~\cite{Hyodo:2007jq,Dote:2008in,Dote:2008hw}. Experimental studies are also
ongoing, in order to extract the unconventional nature of the $\Lambda(1405)$ 
resonance~\cite{Niiyama:2008rt}. In this article, we present a series of recent 
theoretical works on the structure of the $\Lambda(1405)$ resonance in chiral 
dynamics.

\section{FRAMEWORK OF CHIRAL DYNAMICS}

Based on the N/D method~\cite{Oller:2000fj}, we write the general amplitude of 
$s$-wave meson-baryon scattering at total energy $\sqrt{s}$:
\begin{align}
    T(\sqrt{s})
    =& \frac{1}{V^{-1}(\sqrt{s})-G(\sqrt{s};a)} ,
    \label{eq:TChU}
\end{align}
where $V(\sqrt{s})$ is the kernel interaction constrained by chiral symmetry, 
which is given in the matrix form with coupled-channel indices. The function 
$G(\sqrt{s};a)$ is the diagonal matrix with once subtracted dispersion integrals 
of the two-body phase-space function. For the $s$-wave two-body scattering,
$G(\sqrt{s};a)$ contains one subtraction constant $a$ for each channel. 
Identifying the dispersion integral $G(\sqrt{s};a)$ as the loop function with
dimensional regularization, Eq.~\eqref{eq:TChU} is regarded as the solution 
of the algebraic Bethe-Salpeter equation and the subtraction constant $a$ plays a
similar role with the cutoff parameter in the loop function. The interaction 
kernel $V(\sqrt{s})$ is determined by the order by order matching with the 
amplitude in chiral perturbation theory. At leading order, $V(\sqrt{s})$ is given
by the Weinberg-Tomozawa (WT) term
\begin{align}
    V_{ij}(\sqrt{s})
    =-\frac{C_{ij}}{4f^2}
    (2\sqrt{s}-M_i-M_j)
    \label{eq:WTterm} ,
\end{align}
where $C_{ij}$, $M_i$ and $f$ are the group theoretical factor, the mass of the 
baryon in channel $i$, and 
the meson decay constant, respectively. With the leading order WT term, this 
framework is almost equivalent to the old coupled-channel works with vector meson
exchange potential~\cite{Dalitz:1967fp}. Based on chiral perturbation theory, it 
is now possible to include higher order corrections
systematically~\cite{Kaiser:1995eg,Lutz:2001yb}.

This approach has been successfully applied to the $S=-1$ and $I=0$ meson-baryon 
scattering, in which the $\Lambda(1405)$ resonance appears below the $\bar{K}N$ 
threshold. Total cross sections of the $K^-p$ scattering in different final 
states, threshold branching ratios, and the mass spectrum of the $\Lambda(1405)$ 
resonance are well 
reproduced~\cite{Kaiser:1995eg,Oset:1998it,Oller:2000fj,Lutz:2001yb}. Thanks to 
the universal properties of the Weinberg-Tomozawa theorem, the framework of 
chiral dynamics can be applied to many different hadron scattering systems, using
the same mathematical machinery~\cite{Hyodo:2006yk,Hyodo:2006kg}. The successful 
applications to many resonances indicates the power of the chiral dynamics for the 
description of the hadron scattering.

\section{STRUCTURE OF BARYON RESONANCE}

\subsection{Origin of the resonance in the natural renormalization scheme}

In chiral dynamics, the excited baryons are described as resonances in the 
meson-baryon scattering amplitude. In this case, any components beyond the model 
space of two-body states (meson-baryon molecule) are expressed by the 
Castillejo-Dalitz-Dyson (CDD) pole contribution~\cite{Castillejo:1956ed}.
Therefore, to know the importance of the dynamical component, we should extract 
the CDD pole contribution in a transparent manner.

In Ref.~\cite{Hyodo:2008xr}, it is shown that the loop function $G$ can contain 
the CDD pole contribution, in addition to those in the interaction kernel $V$. In
order to visualize all the CDD pole contribution in the kernel $V$, the natural 
renormalization scheme has been proposed. In this renormalization scheme, the 
subtraction constant is determined by purely theoretical argument, excluding the 
CDD pole contribution from the loop function. It is demonstrated that the 
deviation of the subtraction constant from the natural value can be expressed by 
the pole term in the effective interaction kernel. This pole term is interpreted 
as the seed of the resonance in the full amplitude. In this way, the natural 
renormalization scheme, together with the phenomenological amplitude, enables us 
to investigate the origin of the resonances in chiral dynamics. 

Using this technique, we analyze the $\Lambda(1405)$ resonance in the $\bar{K}N$ 
scattering and the $N(1535)$ resonance in the $\pi N$ scattering. It is found 
that the $\Lambda(1405)$ resonance can be constructed mostly by the dynamical 
component, while the $N(1535)$ resonance require some CDD pole contribution on 
top of the dynamical meson-baryon components.

\subsection{Scaling analysis of number of color}

A novel method to investigate the quark structure of the resonances in chiral 
dynamics has been developed by using the $N_c$ scaling for the mesonic 
resonances~\cite{Pelaez:2003dy}. Introducing the $N_c$ dependence through the 
particle masses and the coupling constants, the response of the resonance pole 
position against the variation of $N_c$ can be compared with the general $N_c$ 
scaling low of the mass and width of a $\bar{q}q$ meson, in order to estimate the
size of the $\bar{q}q$ component in the resonance. 

In Refs.~\cite{Hyodo:2007np,Roca:2008kr}, the baron resonances $\Lambda(1405)$ 
and $\Lambda(1670)$ have been studied in the same method. One non-trivial issue 
in the baryonic sector is the $N_c$ dependence of the coupling strength $C_{ij}$ 
in the leading order WT term~\cite{Hyodo:2006yk,Hyodo:2006kg}, which leads to the
existence of the bound state in the large $N_c$ limit. For other parameters such 
as hadron masses and meson decay constant, we adopt the standard $N_c$ dependence
derived from the general argument. In this way we obtain the meson-baryon 
scattering amplitude as a function of $N_c$.

General $N_c$ scaling for the excited baryon $R$ with $N_c$ quarks is given as 
$M_R=\mathcal{O}(N_c)$ and $\Gamma_R=\mathcal{O}(1)$. We calculate the pole 
positions corresponding to the $\Lambda(1405)$ and $\Lambda(1670)$ with different
values of $N_c$. When we change the $N_c$ from $3$ to $12$, the imaginary parts 
of the pole positions for the $\Lambda(1405)$ and $\Lambda(1670)$ change 
drastically, in contrast to the scaling low of $q^{N_c}$ state, 
$\Gamma_R=\mathcal{O}(1)$. This result indicates that the three-quark component 
(at $N_c=3$) in the $\Lambda(1405)$ and $\Lambda(1670)$ should be small.

\subsection{Evaluation of the electromagnetic size}

A standard method to investigate the structure of a particle is to use the 
(virtual) photon probe. The form factor contains the information of the 
electromagnetic structure of the particle. In chiral dynamics, the resonance form
factor can be calculated by introducing an external photon field. Since the 
resonance is expressed microscopically by the bubble sum of the meson-baryon 
loops, the photon field should be attached to the constituent mesons, baryons, 
and vertices.

In Ref.~\cite{Sekihara:2008qk}, the electromagnetic mean squared radii of the 
$\Lambda(1405)$ resonance is studied in chiral dynamics. Result of the charge 
radii of the $\Lambda(1405)$ shows a larger value than the ground state neutron. 
This implies that the charge distribution of the $\Lambda(1405)$ is much wider 
than the neutron in spatial size. This is consistent with the meson-baryon
molecule picture, rather than the compact three-quark structure.

\section{SUMMARY}

We have discussed that the chiral dynamics describes the baryon resonances in 
meson-baryon scattering quite well, and the investigation of the structure of the
resonance is now becoming available. By taking an example of the $\Lambda(1405)$,
we show three approaches to reveal the structure of the resonance: the natural 
renormalization scheme, the $N_c$ scaling method, and the evaluation of the 
electromagnetic size. In view of these three independent analyses, we conclude 
that the $\Lambda(1405)$ resonance is mostly dominated by the meson-baryon 
dynamical component. 

This conclusion may sound as a trivial result by the construction of the model, 
but it is not always the case. For instance, it has been found that the $N(1535)$
resonance and the $\rho$ meson contain substantial contributions of the 
non-dynamical components, although they appear as poles in the scattering 
amplitude, in the same way with the $\Lambda(1405)$. Therefore, the dynamical 
nature for the $\Lambda(1405)$ is a nontrivial consequence in chiral dynamics. On
the other hand, the present analyses clarify the dominant structure only in a 
qualitative manner. More quantitative decomposition of several contributions will
be highly appreciated. At the same time, the analysis of the three-quark like 
resonance, such as $\Delta(1232)$, provides a useful test of the method for the 
structure study. The study of the nontrivial structure of hadron resonances along
this line will shed new light on the nonperturbative dynamics of low energy QCD.

\vspace{0.8cm}
\noindent
{\bf ACKNOWLEDGEMENTS}\\
The author is grateful to Atsushi Hosaka, Daisuke Jido, Luis Roca, and Takayasu 
Sekihara for fruitful collaborations. He thanks Wolfram Weise for kind invitation
to the wonderful symposium. This work is partly supported by the Global Center of
Excellence Program by MEXT, Japan through the Nanoscience and Quantum Physics
Project of the Tokyo Institute of Technology. \\
\noindent

%

\end{document}